\documentclass[aps,prb,twocolumn,showpacs,superscriptaddress]{revtex4} 
\usepackage{graphicx}
\usepackage{amsmath}
\usepackage{color}
\newcommand{\CoChain}{Co\textsuperscript{II}$_4$(OH)$_2$(C$_1$$_0$H$_1$$_6$O$_4$)$_3$}

\begin{document}

\title{Magnetic structure and dynamics of a strongly one-dimensional

 cobalt\textsuperscript{II} metal-organic framework}
\author{Romain Sibille} 
\email[]{rom.sibille@gmail.com}
\affiliation{Institut Jean Lamour, CNRS \& Universit\'e de Lorraine, BP 70239, 54506 Vandoeuvre-l\`es-Nancy, France}
\affiliation{Laboratory for Developments and Methods, Paul Scherrer Institut, 5232 Villigen PSI, Switzerland}
\author{Elsa Lhotel}
\affiliation{Institut N\'eel, CNRS \& Universit\'e Joseph Fourier, BP 166, 38042 Grenoble Cedex 9, France}
\author{Thomas Mazet} 
\affiliation{Institut Jean Lamour, CNRS \& Universit\'e de Lorraine, BP 70239, 54506 Vandoeuvre-l\`es-Nancy, France}
\author{Bernard Malaman}
\affiliation{Institut Jean Lamour, CNRS \& Universit\'e de Lorraine, BP 70239, 54506 Vandoeuvre-l\`es-Nancy, France}
\author{Clemens Ritter}
\affiliation{Institut Laue Langevin, 6 rue Jules Horowitz, 38042 Grenoble, France}
\author{Voraksmy Ban} 
\affiliation{Institute of Condensed Matter and Nanosciences, Universit\'e catholique de Louvain, Place L. Pasteur 1, 1348 Louvain-la-Neuve, Belgium}
\author{Michel Fran\c cois}
\affiliation{Institut Jean Lamour, CNRS \& Universit\'e de Lorraine, BP 70239, 54506 Vandoeuvre-l\`es-Nancy, France}

\begin{abstract}
We investigate the magnetism of the {\CoChain} metal-organic framework which displays complex inorganic chains separated from each other by distances of 1 to 2 nm, and which orders at 5.4~K. The zero-field magnetic structure is determined using neutron powder diffraction: it is mainly antiferromagnetic but posseses a ferromagnetic component along the \textbf{c}-axis. This magnetic structure persists in presence of a magnetic field. Ac susceptibility measurements confirm the existence of a single thermally activated regime over 7 decades in frequency (E/k$_\textsl{B}\approx 64$ K) whereas time-dependent relaxation of the magnetization after saturation in an external field leads to a two times smaller energy barrier. These experiments probe the slow dynamics of domain walls within the chains: we propose that the ac measurements are sensitive to the motion of existing domain walls within the chains, while the magnetization measurements are governed by the creation of domain walls. 
\end{abstract}

\pacs{75.50.Xx, 75.25.-j, 81.07.Pr}
\maketitle
\section{Introduction}
Metal-organic frameworks (MOFs) are organic-inorganic hybrid solids in which a usually low-dimensional inorganic subnetwork is connected with adequately functionalized organic molecules to form a 3D framework. The fascinating structures and properties of these materials have been the focus of intense research efforts over the last 15 years or so.\cite{Cheetham06,Long09,Ferey08} At present, a wide variety of new compounds continues to be discovered, because of the huge combinatorial possibilities offered by the amalgamation of organic and inorganic chemistry. This new class of compounds has even proved able to produce interesting properties for industrial applications.\cite{Czaja09} One of the main reasons for this success is their porous character.\cite{Yaghi99,Ferey05} Various applications have been envisaged for MOFs, principally in the field of gas storage and separation,\cite{Murray09,Li09} but also for catalysis,\cite{Lee09} luminescence,\cite{Allendorf09} or magnetorefrigeration.\cite{Sibille12b} 

MOFs containing magnetic ions furthermore represent a goldmine for low-dimensional and frustrated magnetism.\cite{Kurmoo09} First, when the organic moiety of these materials is composed of magnetically inert molecules such as sufficiently long alkane chains, it can act as a simple spacer that prevents the establishment of electronically-driven magnetic interactions between e.g. spin chains or spin layers. In other words, such systems have a strongly low-dimensional character regarding through-bond magnetic interactions and, consequently, they only become magnetically ordered at low or very-low temperatures, when the one-dimensional (1D) or two-dimensional (2D) correlation length diverges and promotes sizable classical dipolar interactions between correlated spin blocks. Second, in addition to the intrinsic low-dimensionality, magnetic MOFs also show a large diversity in the type of topologies formed by the inter-connection of the magnetic ions (M) through the possible superexchange pathways (M-O-M pathways in most cases). Despite of the clear interest of these materials in the field of magnetism, hybrid frameworks are relatively little studied by neutron scattering techniques,\cite{Feyerherm02,Feyerherm02b,Manson02,Feyerherm03, Mole09, Saines10,Mesbah10,Mesbah10b,Lhotel07,Sibille12d,Mole11,Saines11b,Fabelo11,Fabelo13,Nenert13,Ressouche93,Zheludev05} although these methods are commonly used to investigate the magnetic structures and excitations of purely inorganic compounds. This is partly due to the difficulty in synthesizing deuterated compounds when complex organic ligands are involved, and partly due to the weak magnetic contributions to the total neutron scattering from these materials.

Recently, we reported a new cobalt\textsuperscript{II} hydroxy-sebacate framework of formula {\CoChain}.\cite{Sibille12} The compound contains Co\textsuperscript{II}-based magnetic chains which are very well separated from each other (1 to 2 nm). We revealed the coexistence of magnetic long-range order (LRO) and slow dynamics of the magnetization below $T_t=5.4$ K. This new compound is an excellent candidate for a detailed magnetic study for at least two reasons. First, the magnetic chains are much better isolated from each other than in most of the spin-chain compounds. Second, the chains themselves present a rather complex topology of magnetic interactions between spin carriers which are well known for their large single-ion anisotropy.

Spin-chains attract much attention because they provide genuine examples for testing models that are less tractable in higher dimensions.\cite{Georges02} For example, spin chains with nearest-neighbor antiferromagnetic interactions display a large variety of ground states with non-classical magnetic phenomena.\cite{Tonegawa02,Haldane80,Haldane82} Although ideal 1D spin systems do not undergo LRO at finite temperature due to strong quantum fluctuations,\cite{Bethe31} slight perturbations such as, for instance, weak interchain interactions can make the quantum critical state unstable to three-dimensional (3D) LRO.\cite{Gomez90,Uchiyama99} Another striking feature of 1D magnetic systems is the so-called Single-Chain Magnet (SCM) behavior observed in a relatively large number of metallo-organic compounds.\cite{Caneschi01,Coulon06,Gatteschi13} These systems, though not long-range ordered, present bulk magnetic behavior at low temperature that results from slow intrachain dynamics arising from the combination of a large magnetic anisotropy with ferromagnetic interactions. This was predicted in the pioneering work of R. J. Glauber in 1963.\cite{Glauber63}

In this article we provide deeper insight on the magnetism in the ordered phase of {\CoChain}. In section II we recall the necessary information concerning the crystal structure and macroscopic magnetic behavior of the system at $T\geq2$ K. Experimental details are given in Section III. Our new results are presented and discussed in Section IV. Firstly, the zero-field magnetic structure is determined from neutron powder diffraction data. The magnetic field-dependence of this spin arrangement is qualitatively evaluated. Secondly, we present new magnetic measurements for temperatures down to 70 mK of both magnetization and ac susceptibility, and deduce the characteristic energies of the observed dynamics. Finally, in Section V, we discuss the magnetic structure in relation with the magnetic couplings and anisotropy in the compound. We further propose a scenario to describe the dynamics. Finally, in section VI we summarize the present work.

\section{Previous results}
\subsection{Crystal structure}
\label{struct}

\begin{figure}[t]
\centerline{\includegraphics[width=7.0cm]{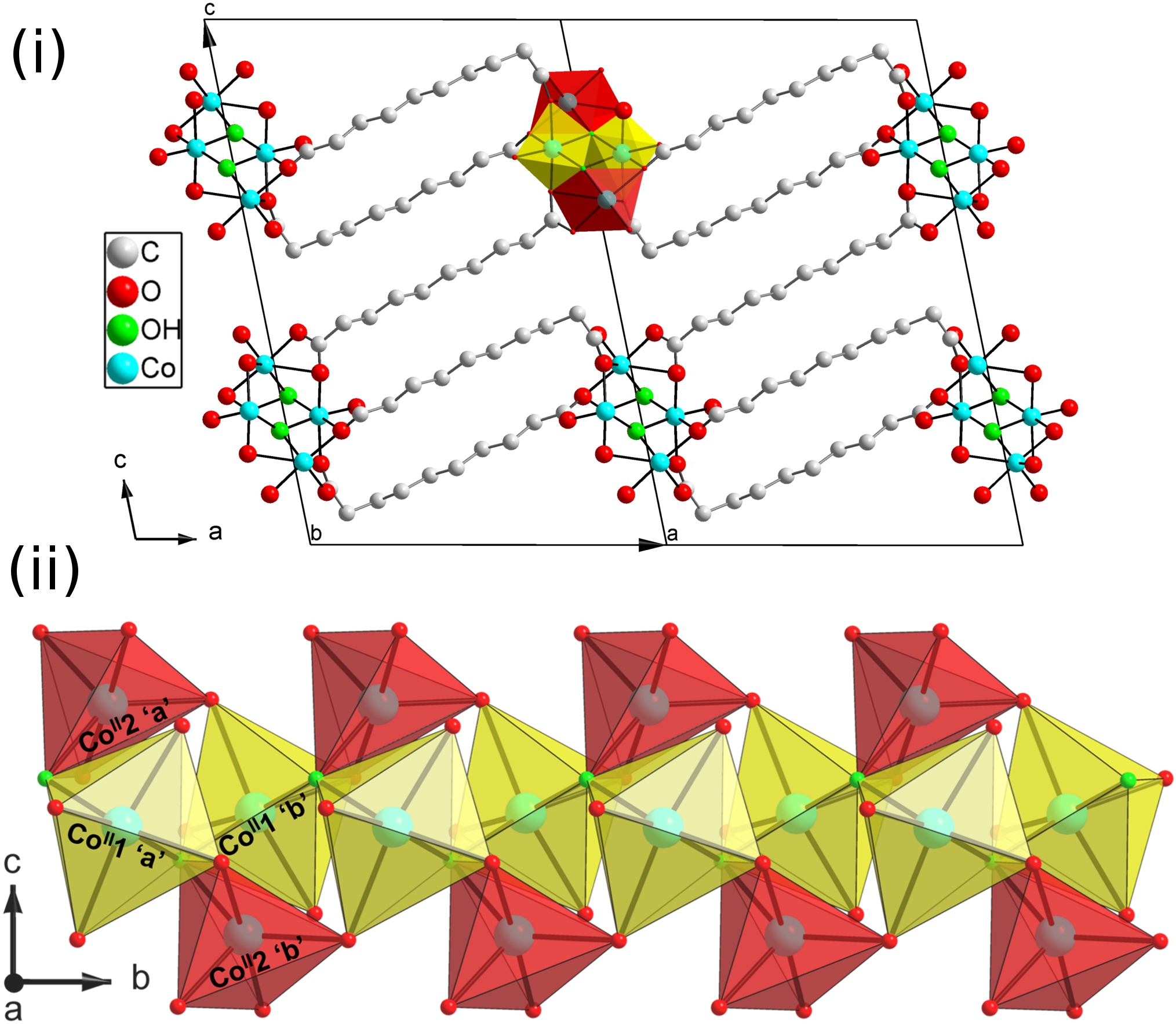}}
\caption{(color online) Crystal structure of {\CoChain}.\cite{Sibille12} Co\textsuperscript{II}1O$_6$ and Co\textsuperscript{II}2O$_5$ polyhedra are represented in yellow and red, respectively. (i) Projection of the 3D metal-organic framework perpendicular to the magnetic chain axis. (ii) Atomic structure of the 1D inorganic subnetwork.}
\label{Fig.1}
\end{figure}

\begin{figure}[t]
\centerline{\includegraphics[width=8cm]{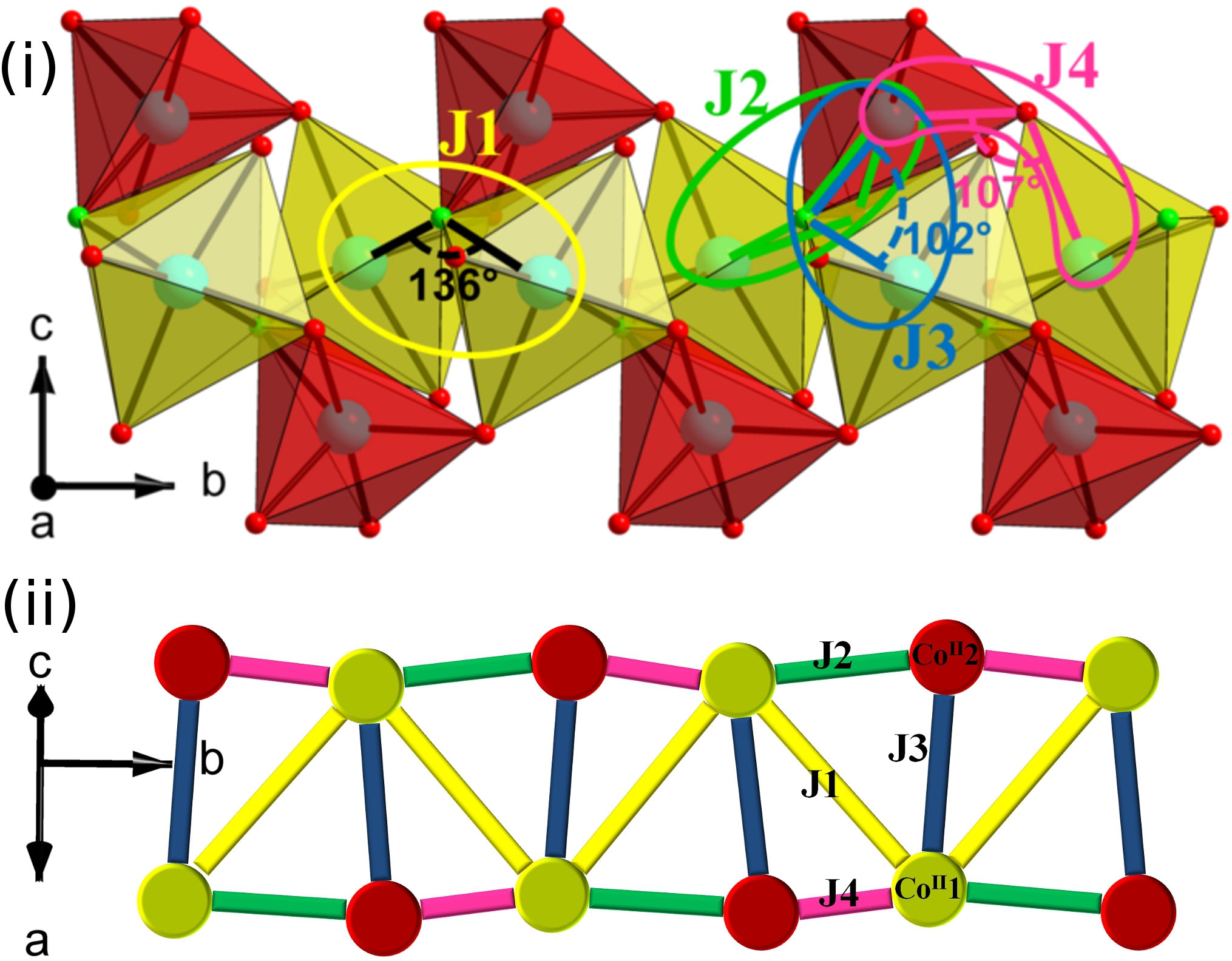}}
\caption{(color online) Intrachain magnetic superexchange pathways. Color codes are the same as in Figure \ref{Fig.1}. (i) J1, J3 and J4 correspond to corner-sharing pathways having the angle values given on the figure, while J2 corresponds to an edge-sharing pathway with Co\textsuperscript{II}1-O-Co\textsuperscript{II}2 angles of about 91\textsuperscript{o} and 96\textsuperscript{o}. (ii) Spin topology viewed along the [101] crystallographic direction, i.e. by applying a 45\textsuperscript{o} rotation around the chain axis from the orientation of Figure \ref{Fig.2}(i).}
\label{Fig.2}
\end{figure}

{\CoChain} adopts a \textsl{P}2$_1$\textsl{/c} monoclinic structure in which magnetic chains develop along the \textbf{b}-axis. Isoreticular compounds, i.e. having a similar framework organization though eventually based on different organic ligands, have been reported with manganese\textsuperscript{II} and iron\textsuperscript{II} ions \cite{Saines11,Sibille12c}. Figure \ref{Fig.1} sheds light on the main structural characteristic of this compound: the strongly 1D character due to large distances between the inorganic chains. The distance between the centers of neighboring chains varies from 11 to 21 \AA, with the shortest path (i.e. in the [001] direction, see Figure \ref{Fig.1}(i)) not involving chemical connections. From the point of view of the magnetism, the combination of these large interchain distances with either the absence of interchain chemical connection or its presence through magnetically inert molecules (alkane chain with 9 consecutive C-C single bonds) precludes any role of interchain interactions other than classical dipolar forces. The 1D inorganic subnetwork of this material contains two crystallographically distinct magnetic ions, Co\textsuperscript{II}1 and Co\textsuperscript{II}2, that have distorted octahedral and distorted trigonal bipyramidal oxygen environments, respectively. Both sites have intrinsically asymmetric crystal field environments as the oxygen atoms originate from carboxylate functions and hydroxide groups. The chains can be initially viewed as being constituted of a central `zig-zag' chain of corner-sharing Co\textsuperscript{II}1O$_6$ octahedra. In addition, Co\textsuperscript{II}2O$_5$ trigonal bipyramids share oxygen atoms with three consecutive Co\textsuperscript{II}1O$_6$ octahedra of the main chain. Both magnetic sites are in the general 4\textsl{e} position which, within a single inorganic chain, splits into two equivalent positions related by the 2-fold screw axis and labeled as `a' and `b' on Figure \ref{Fig.1}(ii).

Four intrachain superexchange pathways can be distinguished and are emphasized on Figure \ref{Fig.2}(i). First, yellow Co\textsuperscript{II}1O$_6$ octahedra are connected to each other along the chain axis by sharing corners through a large superexchange angle of about 136\textsuperscript{o} (J1). Second, each Co\textsuperscript{II}2O$_5$ bipyramid (red) connects with two Co\textsuperscript{II}1O$_6$ octahedra by corner-sharing through smaller superexchange angles ($\sim$102\textsuperscript{o} for J3 and $\sim$107\textsuperscript{o} for J4); and with one Co\textsuperscript{II}1O$_6$ octahedron by edge-sharing (J2) with the nearly orthogonal angles ($\sim$91\textsuperscript{o} and $\sim$97\textsuperscript{o}). The resulting spin topology is represented on Figure \ref{Fig.2}(ii). In addition to these four M-O-M pathways, weaker intrachain magnetic `super-superexchange' pathways exist through the carboxylic functions of the sebacate ligand (M-O-C-O-M pathways), leading to further terms expected to contribute to the full Hamiltonian.

\subsection{Magnetic behavior at T $\geq$ 2 K}
The first magnetic characterizations of {\CoChain} have been previously reported.\cite{Sibille12} The compound undergoes magnetic long-range ordering below T$_t$=5.4 K, as evidenced by heat capacity (C$_p$) measurements and by the appearance of magnetic scattering below T$_t$ as observed by neutron diffraction. The presence of a broad bump in the zero-field C$_p$(T) curve, between the $\lambda$-peak at T$_t$ and 2 to 3 times T$_t$, was attributed to the effects of magnetic correlations within the chains. 

A complex behavior of the magnetic susceptibility was observed above the LRO transition. The non-monotonic shape of the product $\chi T$ could be described by introducing two effective energy parameters (20 and -30 K) which account for both anistropy and (ferromagnetic and antiferromagnetic) exchange interactions. 

Ac susceptibility measurements showed the existence of dynamical processes at T $\leq$ T$_t$. Ac susceptibility data were taken on a Physical Properties Measurement System (Quantum Design, USA) for frequencies of the ac magnetic field between 10 and 6500 Hz. The analysis revealed the existence of a thermally activated behavior within the 3-4.5 K temperature range, with an energy barrier of about 67 K and a characteristic time of the order of 1.4$\times10^{-11}$ s. The observed slow dynamics was attributed to the motion of thermally activated domain walls within the magnetic chains. Finally, ac susceptibility measurements performed under various bias dc magnetic fields revealed a strong displacement of the main peak towards higher temperatures, for both in-phase and out-of-phase components. This latter observation was suggested to evidence the existence of field-induced magnetic order for temperatures above the zero-field transition temperature T$_t$.

\section{Experimental}
\label{exp}
The two powder samples of {\CoChain} used in this work were synthesized hydrothermally according to the method reported previously.\cite{Sibille12} Chemicals (sebacic acid and NaOH) were not deuterated, obtained from commercial suppliers, and used as received. High-purity distilled water was used for the reaction medium. Long-exposure laboratory X-ray powder diffraction revealed the samples to be single phase.

Neutron scattering experiments were conducted using the high-flux D20 two-axis diffractometer at the ILL (Institut Laue Langevin, Grenoble, France). The neutron wavelength was $\lambda = 2.41$ \AA. All experiments were performed with a 6 mm diameter vanadium sample container and a cryomagnet providing temperatures down to 1.8 K and magnetic fields up to 6 T. Zero-field data were first recorded by simply loading the 2 g powder into the sample holder. Two long duration neutron diffraction patterns were recorded at 1.8 K and 20 K (8 hours for each). 

For measurements with an applied magnetic field, the sample was immersed in deuterated isopropanol in order to avoid reorientation of the powder particles. Isopropanol freezes around T $\sim$ 115 K in a glassy state, giving rise to a weak and broad bump in the data around $2\theta = 35\textsuperscript{o}$. After initial rapid cooling of the sample in the remnant field of the magnet, the temperature was never raised above the glass transition temperature of isopropanol for the remainder of the experiment. It was therefore guaranteed that the powder grains remained fixed and that the same powder average was measured in each run. A diffraction pattern was recorded under these conditions, in zero-field at 10 K, and used for subtraction of the nuclear scattering. Data were then recorded under zero-field and under two different applied magnetic fields (0.5 and 2 T). For each of these three field values, diffraction data were collected for different temperatures across the transition, from 1.8 K to higher temperatures. After these measurements, another zero-field 10 K pattern was recorded in order to confirm the absence of reorientation of the powder particles. 

The neutron patterns were analyzed by Rietveld refinements using the Fullprof program. \cite{Carvajal93} The shape of the diffraction peaks was fitted using a pseudo-Voigt function. Besides the six peak profile parameters, the fit of the paramagnetic pattern recorded at 20 K comprised zero shift, cell parameters, preferred orientation of the powder particles, and scale factor. The agreement indices R$_{wp}$ and R$_F$ are the factors defined in Ref. \onlinecite{McCusker99}. The difference resulting from the two long-duration patterns recorded at 1.8 and 20 K was used for the refinement of the magnetic structure. All parameters were fixed to their previously refined values (20 K), except the lattice parameters which were taken from a Le Bail decomposition\cite{LeBail88} of the 1.8 K pattern. The agreement factor noted R$_{magn}$ is the Bragg agreement factor\cite{McCusker99} for the fit of the magnetic scattering observed on the difference pattern. The representational analysis techniques described by Bertaut\cite{Bertaut63} were used to determine the possible magnetic structures compatible with both the magnetic wavevector and the symmetry of the crystal structure. The irreducible representations of the propagation vector group were determined with the program BasIreps\cite{Carvajal10}. The possible solutions were first tested by refining the mixing coefficients of the basis vectors, as implemented in the Fullprof program.\cite{Carvajal93} The final Rietveld refinement of the magnetic scattering was performed in the `spherical coordinates procedure' of the software. The three parameters (R, $\theta$, $\varphi$) per Co\textsuperscript{II} characterizing their atomic magnetic moments correspond to the value of the magnetic moment $|M|$ (in Bohr magnetons), the polar $\theta$ angle with z axis and the azimuth $\varphi$ angle between the x axis and the projection of R onto the (x,y) plane. Since this mode works only if z is perpendicular to the (x,y) plane, the present monoclinic space group was described in the Laue Class 1 1 2\textsl{/m} for the magnetic refinement.

Two powder samples (sample 1 - mass $m=1.96$ mg and sample 2 - mass $m=0.4$ mg) were measured using low temperature superconducting quantum interference device (SQUID) magnetometers equipped with a miniature dilution refrigerator developed at the Institut N\'eel-CNRS Grenoble \cite{Paulsen01}. Sample 1 was obtained from the 2 g sample used for neutron scattering experiments. The samples were prepared in a copper pouch and mixed with apiezon grease (for thermalization and to avoid reorientation of the powder under magnetic field). They were attached to a copper tress suspended from the mixing chamber of the dilution fridge. $M(H)$ hysteresis loops were measured for both samples in the `high' field magnetometer which reaches 8 T. Ac and relaxation measurements were performed in the `low' field magnetometer, which reaches 3500 Oe, for sample 1. The magnitude of the alternating field was $H_{ac}=1$ Oe.

\section{Results}
\subsection{Magnetic structure}
\label{mag_struc}

\begin{figure}[t]
\centerline{\includegraphics[width=7.0cm]{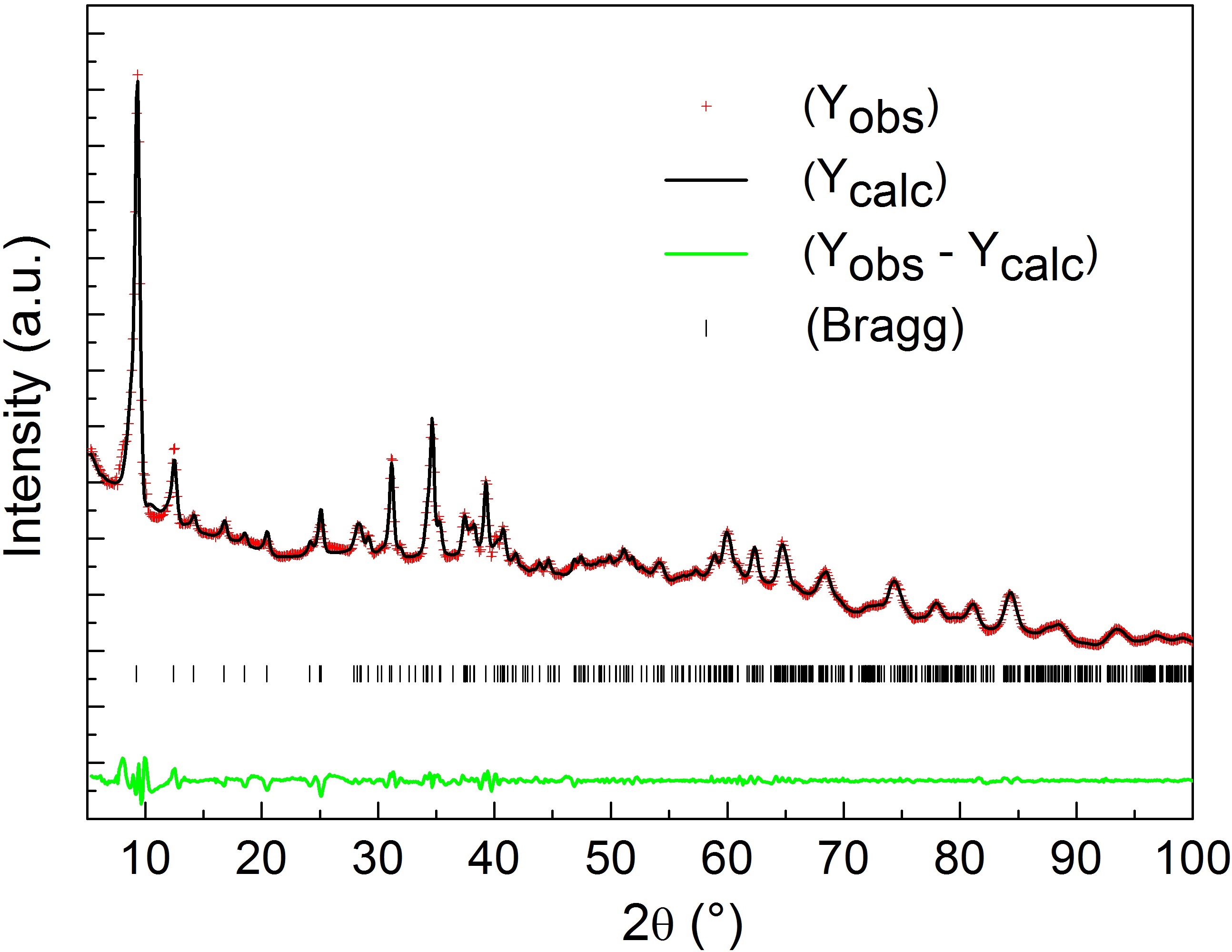}}
\caption{(color online) Rietveld plot of the neutron powder diffraction data at 20 K obtained from the X-ray single-crystal diffraction structural model at 100 K. Two intensity-dependent parameters were refined (scale factor and preferred orientation parameter).}
\label{Rietveld_nuclear}
\end{figure}

The Rietveld refinement of the 20 K neutron pattern was obtained by using the atomic coordinates determined from X-ray single-crystal diffraction at 100 K (see Figure  \ref{Rietveld_nuclear})\cite{Sibille12}. The good agreement between the calculated and recorded neutron patterns (R$_{wp}$ = 0.115 and R$_F$ = 0.127) suggests the absence of significant modifications of the crystal structure below 100 K. The refined unit cell parameters are: $a = 15.16(1)$ \AA, $b = 4.884(3)$ \AA, $c = 22.53(2)$ \AA, and $\beta = 100.15(6)$ \textsuperscript{o}.

The magnetic scattering was extracted from the data recorded at 1.8 K by subtracting the 20 K pattern. The resulting `difference pattern' shows weak magnetic Bragg peaks both on top of some nuclear peaks of the (\textsl{h},0,0) type and on the (0,0,\textsl{l}) and (\textsl{h},0,\textsl{l}) peaks with odd \textsl{l} indices. These two latter families of reflections, absent in the high-temperature pattern, are forbidden by the \textsl{c}-glide of the nuclear space group. These indications for such a non-Bravais lattice (breaking of the nuclear point-group symmetries) imply a commensurate magnetic structure with a propagation vector \textbf{\textsl{k}}=(0,0,0).

To simplify the analysis, group theory was used in order to predict the magnetic arrangements compatible with the nuclear symmetry, thus reducing the number of independent parameters in the refinement. In the Landau theory, magnetic fluctuations in the paramagnetic state while approaching a second-order phase transition reflect the crystal symmetries. At the transition, one magnetic mode is stabilized. The magnetic modes compatible with the nuclear space group \textsl{G} (= \textsl{P}2$_1$\textsl{/c}) are given by the irreducible representations of the little group \textsl{G$_k$} that leave \textbf{\textsl{k}} invariant, i.e. the irreducible representations of \textsl{G} for \textbf{\textsl{k}}=(0,0,0). The application of the group theory, using the irreducible representations tabulated by Kovalev\cite{Kovalev93} and available in the BasIreps program,\cite{Carvajal10} shows that the group representation can be reduced to a sum of four irreducible representations of order 1: $\Gamma=3\tau\textsuperscript{1}\oplus3\tau\textsuperscript{2}\oplus3\tau\textsuperscript{3}\oplus3\tau\textsuperscript{4}$.

\begin{table*}[t]
\begin{ruledtabular}
\begin{tabular}{lcccccccccccccccccc}
& & & & & $\tau\textsuperscript{1}$ & & & & $\tau\textsuperscript{2}$ & & & & $\tau\textsuperscript{3}$ & & & & $\tau\textsuperscript{4}$ & \\ \cline{5-7} \cline{9-11} \cline{13-15} \cline{17-19}
Co\textsuperscript{II}1 or Co\textsuperscript{II}2 & & & & \textsl{m}$_x$ & \textsl{m}$_y$ & \textsl{m}$_z$ & & \textsl{m}$_x$ & \textsl{m}$_y$ & $\textsl{m}_z$ & & $\textsl{m}_x$ & $\textsl{m}_y$ & $\textsl{m}_z$ & & $\textsl{m}_x$ & $\textsl{m}_y$ & $\textsl{m}_z$ \\ \hline
Position 1& & \textbf{g$_1$} && + & + & + & & + & + & + & & + & + & + & & + & + & + \\
Position 2& & \textbf{g$_2$} && - & + & - & & - & + & - & & + & - & + & & + & - & + \\
Position 3& & \textbf{g$_3$} && + & + & + & & - & - & - & & + & + & + & & - & - & - \\
Position 4& & \textbf{g$_4$} && - & + & - & & + & - & + & & + & - & + & & - & + & - 
\\
\end{tabular}
\end{ruledtabular}
\caption{\label{Irreps} Magnetic components for the four Co\textsuperscript{II} positions generated for both Co\textsuperscript{II}1 and Co\textsuperscript{II}2 atoms. The four equivalent positions for these two atoms in the general 4\textsl{e} Wyckoff position are generated by the symmetry elements \textbf{g$_1$} = 1, \textbf{g$_2$} = 2$_1$ (0, y, 1/4), \textbf{g$_3$} = $\bar1$ (0, 0, 0), \textbf{g$_4$} = \textsl{c} (x, 1/4, z).}
\end{table*}

\begin{figure}[t]
\centerline{\includegraphics[width=7.5cm]{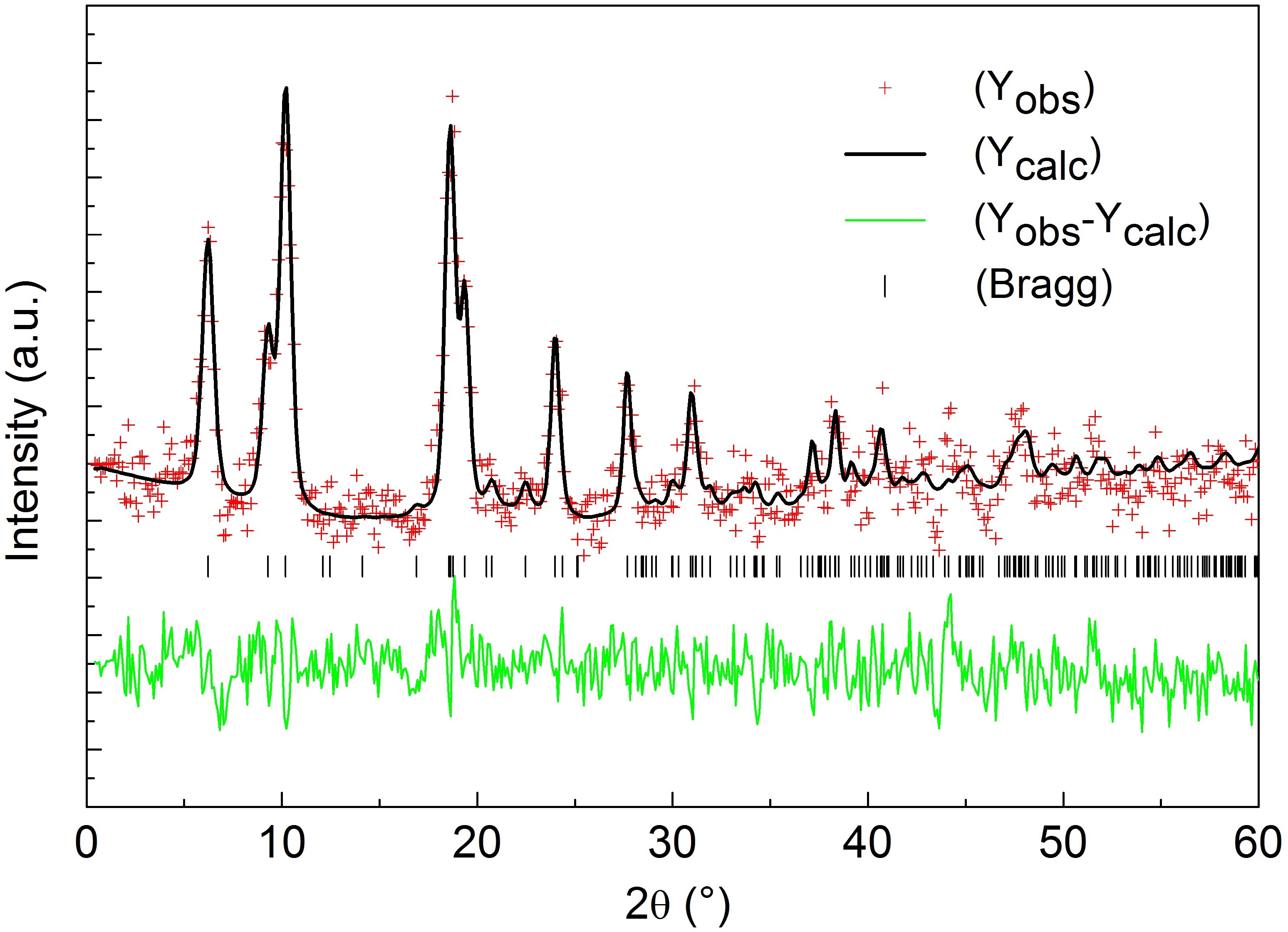}}
\caption{(color online) Rietveld refinement of the magnetic neutron scattering at 1.8 K. The data correspond to the difference between the two long duration patterns recorded at 1.8 K and 20 K. The black line is the fit of the data according to the $\tau\textsuperscript{3}$ model (see text for details).}
\label{Rietveld_magnet}
\end{figure}

The four possible magnetic arrangements resulting from this analysis are summarized in Table \ref{Irreps}.
In principle, $\tau\textsuperscript{2}$ and $\tau\textsuperscript{4}$ can be excluded because they do not contain a ferromagnetic coupling while magnetic scattering is detected on the top of allowed nuclear peaks. Moreover, testing the four models by refining the mixing coefficients of the basis vectors of the two magnetic moments indicates $\tau\textsuperscript{3}$ to most likely describe the magnetic structure. The three other models provide unreasonable magnetic moment values and/or much larger agreement factors. Imposing the irreducible representation $\tau\textsuperscript{3}$, the refinement of the six mixing coefficients converges and gives reasonable magnetic moment values. In particular, the magnetic moment of Co\textsuperscript{II}1 is oriented along the chain axis (\textbf{b}-axis), implying that the Co\textsuperscript{II}1 sublattice forms a simple antiferromagnetic chain. In a second step, the refinement was realized by describing the magnetic moments by their spherical coordinates (see Section \ref{exp}), in order to improve the stability of the refinement procedure. This procedure confirmed the orientation of the Co\textsuperscript{II}1 moment to lie along the \textbf{b}-axis. Thus, we constrained the magnetic moment of Co\textsuperscript{II}1 with $\varphi_1 = 0$\textsuperscript{o} and $\theta_1$ = 0$\textsuperscript{o}$ and 180$\textsuperscript{o}$ for the positions related by the screw axis, in accordance with $\tau\textsuperscript{3}$. The magnetic moment of Co\textsuperscript{II}2, for its part, was refined with its three spherical coordinates by imposing the couplings relevant to $\tau\textsuperscript{3}$: ($|M_2|$, $\varphi_2$, $\theta_2$) for Co$\textsuperscript{II}2$ ions in positions 1 and 3, and ($|M_2|$, $\varphi_2$, $180-\theta_2$) for Co$\textsuperscript{II}2$ in positions 2 and 4. Using these four parameters ($|M_1|$ for Co$\textsuperscript{II}1$ and $|M_2|$, $\varphi_2$, $\theta_2$ for Co$\textsuperscript{II}2$), the refinement procedure converges satisfactorily (R$_{magn}=0.10$, $\chi\textsuperscript{2}$=1.49, see Figure  \ref{Rietveld_magnet}). Starting from this model, lifting the additional constraints applied to the magnetic moment of Co\textsuperscript{II}1 (${\bf M_1}//\textbf{b}$) does not lead to a significantly different magnetic structure. In the final model, the refined $\varphi_2$ value of about $100\textsuperscript{o}$ corresponds to the monoclinic angle of the unit cell, indicating that the Co$\textsuperscript{II}2$ magnetic moment is roughly in the (\textbf{b},\textbf{c}) plane. Consequently, $\theta_2$ ($\sim35\textsuperscript{o}$) defines both the rotation of the Co$\textsuperscript{II}2$ magnetic moments in the (\textbf{b},\textbf{c}) plane relative to the chain axis, and their canting angle between the positions related by the screw axis ($180-2\times\theta\sim110\textsuperscript{o}$).

\begin{figure}[t]
\centerline{\includegraphics[width=8.5cm]{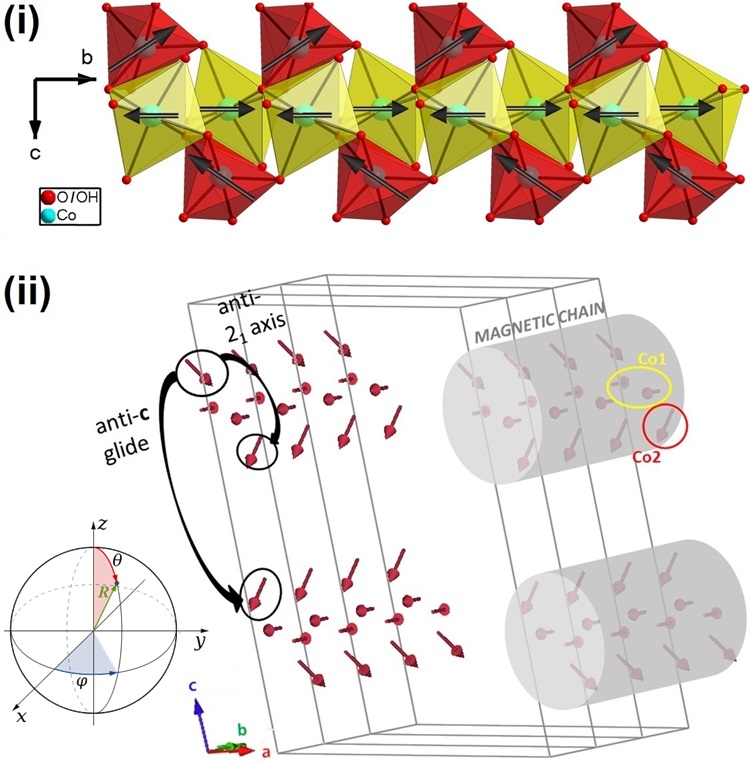}}
\caption{(color online) Zero-field magnetic structure at 1.8 K. (i) View of the magnetic chain along the \textbf{a}-axis showing the orientation of the magnetic moments relative to their chemical environments. (ii) Global view of the magnetic structure. The inset on the left reminds the general definition of spherical angles. The polar and azimuth angles given in the text and in Table \ref{table_magmom} correspond to a setting with x = \textbf{a}, y = \textbf{c}, z = \textbf{b} (Laue Class 1 1 2\textsl{/m}, see Section \ref{exp}).}
\label{magn_struct}
\end{figure}

This refinement gives magnetic moment values of 2.19(7) and 2.68(8) for Co$\textsuperscript{II}1$ and Co$\textsuperscript{II}2$ respectively (see Figure \ref{magn_struct} and Table \ref{table_magmom}). The Co$\textsuperscript{II}1$ sublattice is fully compensated (${\bf M_1}//\textbf{b}$), while Co$\textsuperscript{II}2$ has both compensated ($|\textsl{m}_{y2}|=2.13(5)~\mu_B$) and uncompensated ($|\textsl{m}_{z2}|=1.64(8)~\mu_B$) components. The global ferromagnetic component for this material is thus of about $0.82~\mu_B$.Co$\textsuperscript{II}$mol$\textsuperscript{-1}$. This value, after averaging over the three directions ($0.82\div3\sim0.27~\mu_B$.Co$\textsuperscript{II}$mol$\textsuperscript{-1}$), is in good agreement with the size of the magnetization jump at very low field ($\sim0.4~\mu_B$.Co$\textsuperscript{II}$mol$\textsuperscript{-1}$, see Section \ref{MH} and Ref. \onlinecite{Sibille12}). The magnetic structure is presented on Figure \ref{magn_struct}, and is further discussed in Section \ref{discussion}.

\begin{table}
\begin{ruledtabular}
\begin{tabular}{*{7}{c}}
                            & \textsl{m}$_x$ & \textsl{m}$_y$ & \textsl{m}$_z$ & \bf M     & $\theta$ &$\varphi$    \\ \hline
Co$\textsuperscript{II}$1-1 & 0              & 2.19(7)        & 0              &   2.19(7) & 0       & 0           \\
Co$\textsuperscript{II}$1-2 & 0              & -2.19(7)       & 0              &   2.19(7) & 180     & 0           \\
Co$\textsuperscript{II}$1-3 & 0              & 2.19(7)        & 0              &   2.19(7) & 0       & 0           \\
Co$\textsuperscript{II}$1-4 & 0              & -2.19(7)       & 0              &   2.19(7) & 180     & 0            \\ \hline
Co$\textsuperscript{II}$2-1 & 0              & -2.13(5)       & -1.64(8)       &   2.68(8) & -36(2)  & 101(3)       \\
Co$\textsuperscript{II}$2-2 & 0              & 2.13(5)        & -1.64(8)       &   2.68(8) & 216(2)  & 101(3)       \\
Co$\textsuperscript{II}$2-3 & 0              & -2.13(5)       & -1.64(8)       &   2.68(8) & -36(2)  & 101(3)       \\
Co$\textsuperscript{II}$2-4 & 0              & 2.13(5)        & -1.64(8)       &   2.68(8) & 216(2)  & 101(3)       \\
\end{tabular}
\end{ruledtabular}
\caption{Results from the Rietveld refinement of the 1.8 K magnetic scattering (see Figure \ref{Rietveld_magnet}). The cartesian components of the magnetization along the crystallographic axes and the norm of the axial vector \textbf{M} are given in Bohr magnetons. The spherical angles $\theta$ and $\varphi$ are expressed in degree.}
\label{table_magmom}
\end{table}

\subsection{Neutron diffraction under magnetic field}
\label{neut_field}

\begin{figure}[h]
\centerline{\includegraphics[width=5.5cm]{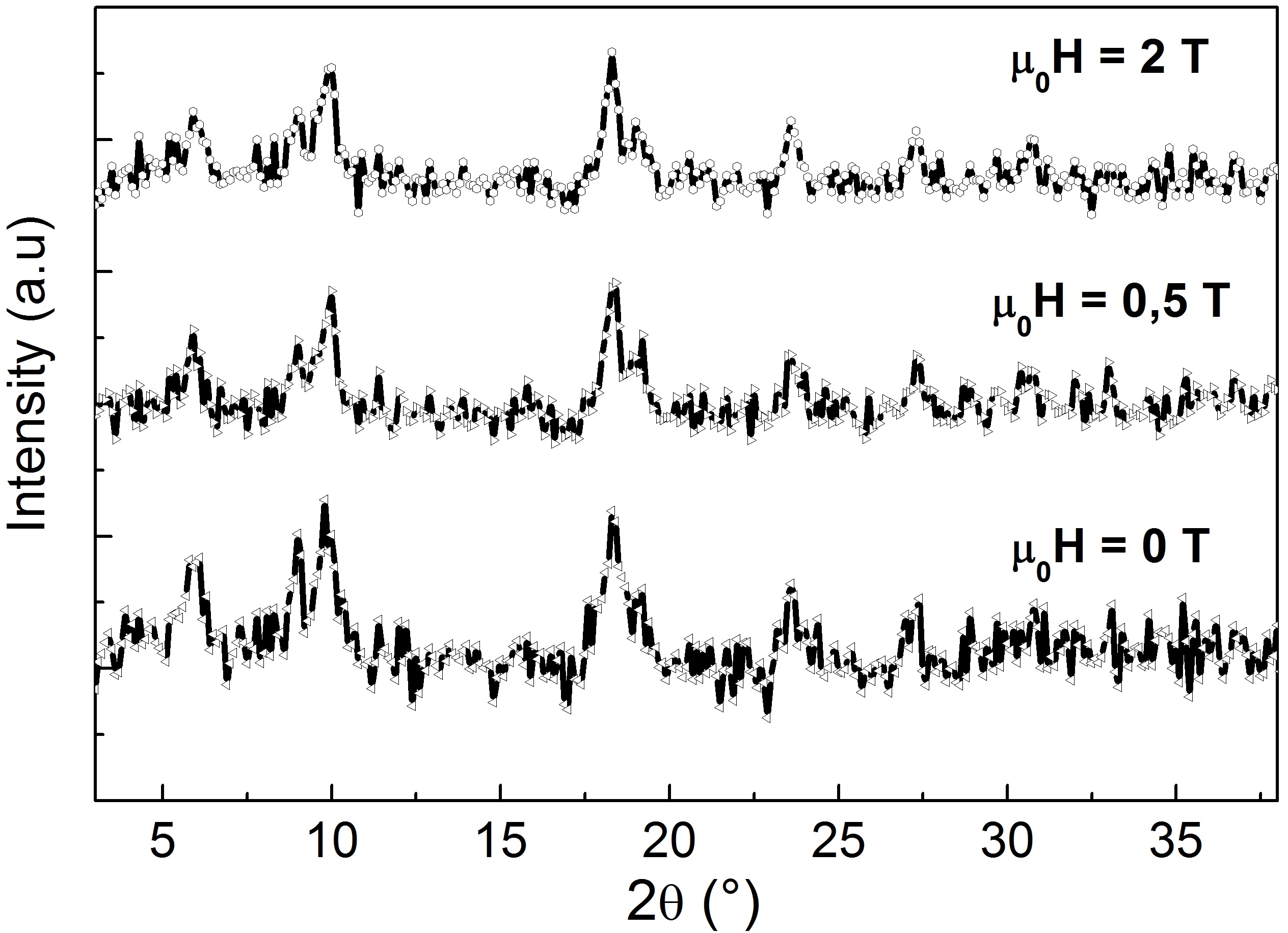}}
\caption{Magnetic scattering at 1.8 K for several applied magnetic fields. For each field, the data correspond to differences between patterns recorded at 1.8 K and at 10 K in zero-field.}
\label{patterns_3H}
\end{figure}

The stability of the magnetic structure against the application of an external magnetic field was verified by measuring the magnetic scattering at 1.8 K under fields of 0.5 and 2 T (see Section \ref{exp}). Despite the complex experimental setup (cryomagnet and dispersion of the compound in deuterated isopropanol), the high hydrogen content of the compound, and the diluted character of the magnetism, weak magnetic elastic scattering could be observed (see Figure \ref{patterns_3H}). Although the bad statistics preclude refinement of these data, the absence of drastic changes in the 1.8 K magnetic structure is evidenced up to 2 T. This result is in agreement with the shape of the isothermal magnetization at 1.8 K (see Section \ref{MH} and Ref. \onlinecite{Sibille12}), i.e. a fast increase at very low field due to the ferromagnetic component of the magnetization, followed by a mild increase due to competition between the applied field and the local anisotropy of Co$\textsuperscript{II}$ ions.

\begin{figure}[h]
\centerline{\includegraphics[width=8.5cm]{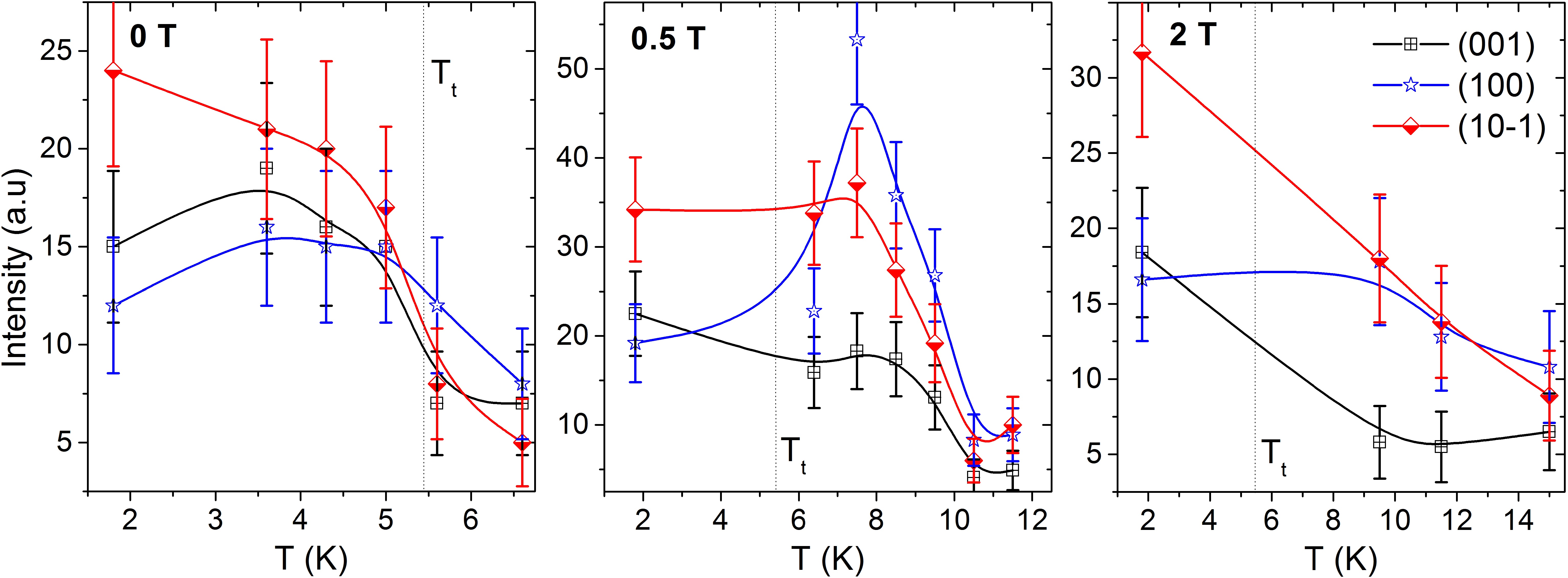}}
\caption{(color online) Maximal intensity at the positions of (001), (100), and (10$\bar1$) reflections recorded upon warming under zero-field (left), 0.5 T (middle), and 2 T (right). The plotted magnetic neutron scattering is obtained after subtraction of a pattern recorded in zero-field at 10 K.}
\label{3ref_H}
\end{figure}

Data were also recorded as a function of temperature for 0, 0.5 and 2 T. Figure \ref{3ref_H} summarizes the obtained data through the thermal evolution of three magnetic Bragg reflections. 
We can first look at the (10$\bar1$) reflection, which is characteristic of the interchain correlations (anti-\textbf{c} reflection, see Figure \ref{magn_struct}). In zero-field (see figure \ref{3ref_H}), the (10$\bar1$) intensity increases below $T_t=5.4$ K, the transition temperature determined from macroscopic measurements \cite{Sibille12}. 
When a magnetic field is aplied, the intensity of this reflection remains finite above $T_t$. This indicates the persistence of interchain correlations in presence of an applied field well above $T_t$: up to 10 and 13 K for applied field values of 0.5 and 2 T respectively. This is not surprising since the interchain magnetic arrangement is ferromagnetic-like, so that a magnetic field is expected to stabilize these correlations. It is worth noting that these results are in good agreement i) with the previously reported phase diagram established from the bias dc field dependence of the ac susceptibility maxima,\cite{Sibille12} and ii) with the position of the maxima in the derivative of the temperature dependence of the magnetization: 9.8~K for 0.5 T and 13.9 K for 2 T (data not shown).\cite{Sibille12c}

Second, we consider the (001) and (100) reflections. These reflections are relevant for observing the evolution of the canting within the chains. 
For instance, considering $\varphi\sim100\textsuperscript{o}$, (001) is maximal for $\theta=0\textsuperscript{o}$ while (100) is completely absent, and vice versa for $\theta=90\textsuperscript{o}$.
The temperature dependence of these reflections depends strongly upon the magnetic field, especially at the onset of the interchain correlations when field-induced magnetic scattering is detected. This indicates significant rearrangements of the spin structure within the chains at temperatures for which magnetic field induces long-range ordering, and points toward a rich $H-T$ phase diagram. 
Interestingly, at low temperature, the respective intensities of these reflections are little affected by the magnetic field up to 2 T. This suggests that the applied magnetic field only provokes a small canting with respect to the zero field magnetic structure. 

\subsection{Hysteresis loops and magnetic avalanches}
\label{MH}
\begin{figure}[t]
\centerline{\includegraphics[width=8cm]{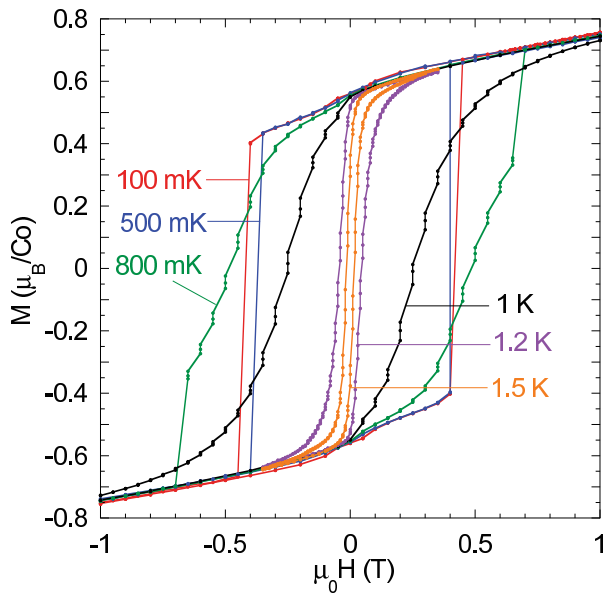}}
\caption{(color online) Hysteresis loops $M$ vs. $H$ for sample 1 at several temperatures between 100 mK and 1.5 K, measured with field increments of 500 Oe every 110 s for $T \le 1$ K, and of 50 Oe every 80 s for $T>1$ K. }
\label{fig_hyst1}
\end{figure}

\begin{figure}[t]
\centerline{\includegraphics[width=8cm]{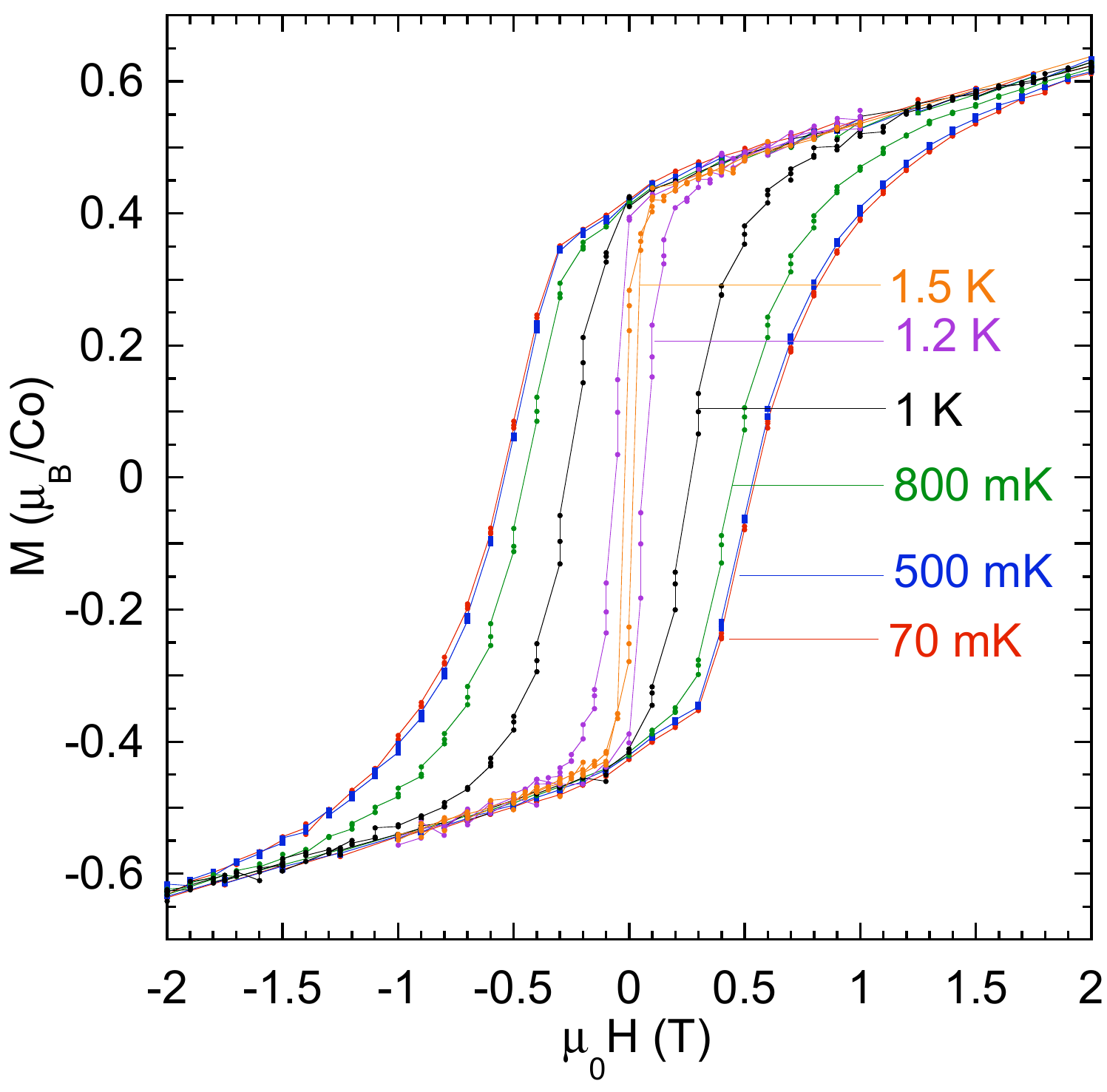}}
\caption{(color online) Hysteresis loops $M$ vs. $H$ for sample 2 at several temperatures between 100 mK and 1.5 K, measured with field increments of 1000 Oe every 110 s for $T \le 1$ K, and of 500 Oe every 110 s for $T>1$ K. }
\label{fig_hyst2}
\end{figure}

$M(H)$ hysteresis loops start to open below 2 K as shown in Figures \ref{fig_hyst1} and \ref{fig_hyst2}. The maximum coercive field is about 0.5 T at low temperature. These hysteresis loops are strongly affected by the dynamics existing in the chains, resulting in relaxation at a given field during the extraction measurements. This can be seen clearly in the data at 800~mK or 1 K, for example.

Below 1 K, magnetic avalanches (that is to say, abrupt reversals of the magnetization) were measured in sample~1, at fields between 0.4 and 0.5 T (see Figure \ref{fig_hyst1}). These avalanches are typical of samples with slow relaxation processes at very low temperature (see for example Ref. \onlinecite{Mn12}) and have been observed in other ordered spin-chains \cite {Lhotel06, Lhotel08}. The avalanche mechanism can be described as follows: due to slow dynamics (large energy barrier, pinned domain-walls, \dots), when the sample is driven out of thermodynamic equilibrium, by an applied field, the magnetization does not follow the field. At a given field (the resonant tunneling field or the nucleation field for example), the reversal of the magnetization becomes easier, and spins can start to flip. Then, because the system is at very low temperature, the specific heat is very low and the energy $\Delta S . H$ released in this reversal heats the neighboring spins, which in turn can flip. This process results in an increase of the internal temperature of the sample, which itself induces a faster relaxation, and provokes a global reversal of the magnetization termed a magnetic avalanche.
In some cases avalanches can be suppressed by slowing the field sweeping rate in the hysteresis loop, or by improving the thermal contact between the sample and the mixing chamber of the dilution fridge which provides the cooling power. The latter was done in the present compound with sample 2, which was better thermalized than sample 1 (smaller size and more apiezon grease). Indeed, magnetic avalanches could be suppressed in this sample (see Figure \ref{fig_hyst2}). Interestingly, around $H_c=0.4$ T, there is however a change of slope in the magnetization curve, showing that the magnetization starts to respond to the applied magnetic field. This indicates that $H_c$ corresponds to an energy scale of the system, and that the previously observed avalanches trigger at a field which is characteristic of a faster relaxation of the system. However, because the measurements were made on a powder and the magnetic structure is very complex, the origin of this energy scale is not readily determinable.

\subsection{Dynamics at very low temperature}
As shown in Ref. \onlinecite{Sibille12}, and observed here in the $M(H)$ hysteresis loops, \CoChain~presents slow dynamics in the ordered regime, characterized by an energy barrier of about 60 K which is attributed to domain-wall dynamics. In order to get a better understanding of the dynamics we have conducted complementary experiments at  temperatures lower than 2 K. 

\begin{figure}[t]
\centerline{\includegraphics[width=8cm]{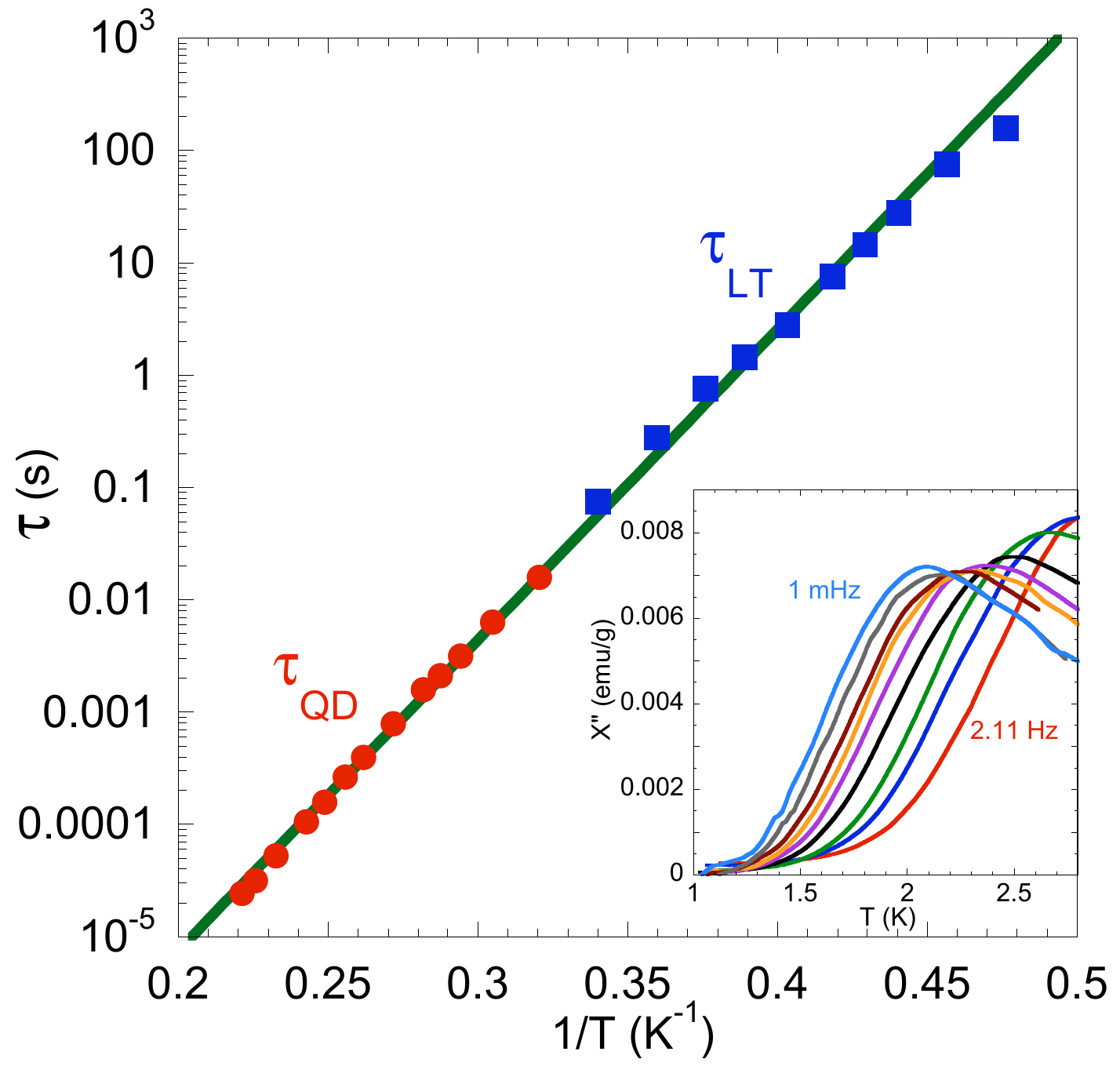}}
\caption{(color online) Relaxation time $\tau=1/2 \pi f$ vs. $1/T_{peak}$ in a semi-logarithmic scale from ac susceptibility measurements. Red points are from MPMS (QD) measurements (10 Hz$<f<$ 6.5 kHz) \cite{Sibille12}, and blue squares from measurements in dilution fridge magnetometers (LT) (1 mHz $<f<2.11$ Hz). The line is a fit to the Arrhenius law with $\tau_0\approx2.2 \times 10^{-11}$ s and $E_b\approx64$ K. Inset: Imaginary part $\chi"$ of the susceptibility at low temperature. }
\label{fig_tau}
\end{figure}

First, we have measured the ac susceptibility from 2.11 Hz down to the lowest frequency reachable in our magnetometer, 1~mHz. In this frequency range, the ac susceptibility still shows a frequency dependence, which we characterize by tracking the maximum in the imaginary part of the susceptibility $\chi"$ (see inset of Figure \ref{fig_tau}). Expecting that, at the maxima, the measurement frequency $f$ matches the relaxation time $\tau$ of the system, we have plotted $\tau=1/2 \pi f$ as a function of the inverse of the temperature of the maximum $T_{peak}$, together with the higher frequency results from Ref. \onlinecite{Sibille12} (see Figure \ref{fig_tau}). These data can be fitted with an Arrhenius law $\tau=\tau_0 \exp (E_b/k_BT)$ with $\tau_0\approx2.2 \times 10^{-11}$~s and $E_b\approx64$~K in the whole frequency range, that is to say, over 7 decades. This shows that the mechanism behind the dynamics is robust in the ordered regime, and remains thermally activated down to at least 2.1 K. The existence of a distribution around the central barrier energy E$_b$ was previously suggested by the shape of the Cole-Cole plots ($\chi"$ vs. $\chi'$).\cite{Sibille12} The parameter extracted from the fit of these plots to a generalized Debye model\cite{Hagiwara98} ($\alpha\approx0.4$ for T = 4 K, see Ref. \onlinecite{Sibille12}), can be used to estimate the full width at half maximum of this distribution ($\Delta E _{1/2}\approx 16$ K), by following the phenomelogical formalism developped by Mydosh for the study of spin glasses.\cite{Mydosh86,Lhotel13}

To probe the dynamics at lower temperatures, wherein we expect a slower response, we have performed measurements of the relaxation of the magnetization as a function of time. The protocol is the following: (i) the sample is heated to about 2 K (temperature at which there is no hysteresis in the magnetization) \cite{footnote1}, (ii) a field $H=-3500$~Oe is applied, (iii) the sample is cooled towards the target temperature, (iv) the field is removed and the magnetization is measured as a function of time. We note that, according to the neutron scattering experiments under applied magnetic field (see Section \ref{mag_struc}), a field of 3500 Oe does not seem to destabilize the magnetic structure. This implies that the induced magnetization is due to the reversal of magnetic domains or to canting phenomena. 

\begin{figure}[t]
\centerline{\includegraphics[width=7.5cm]{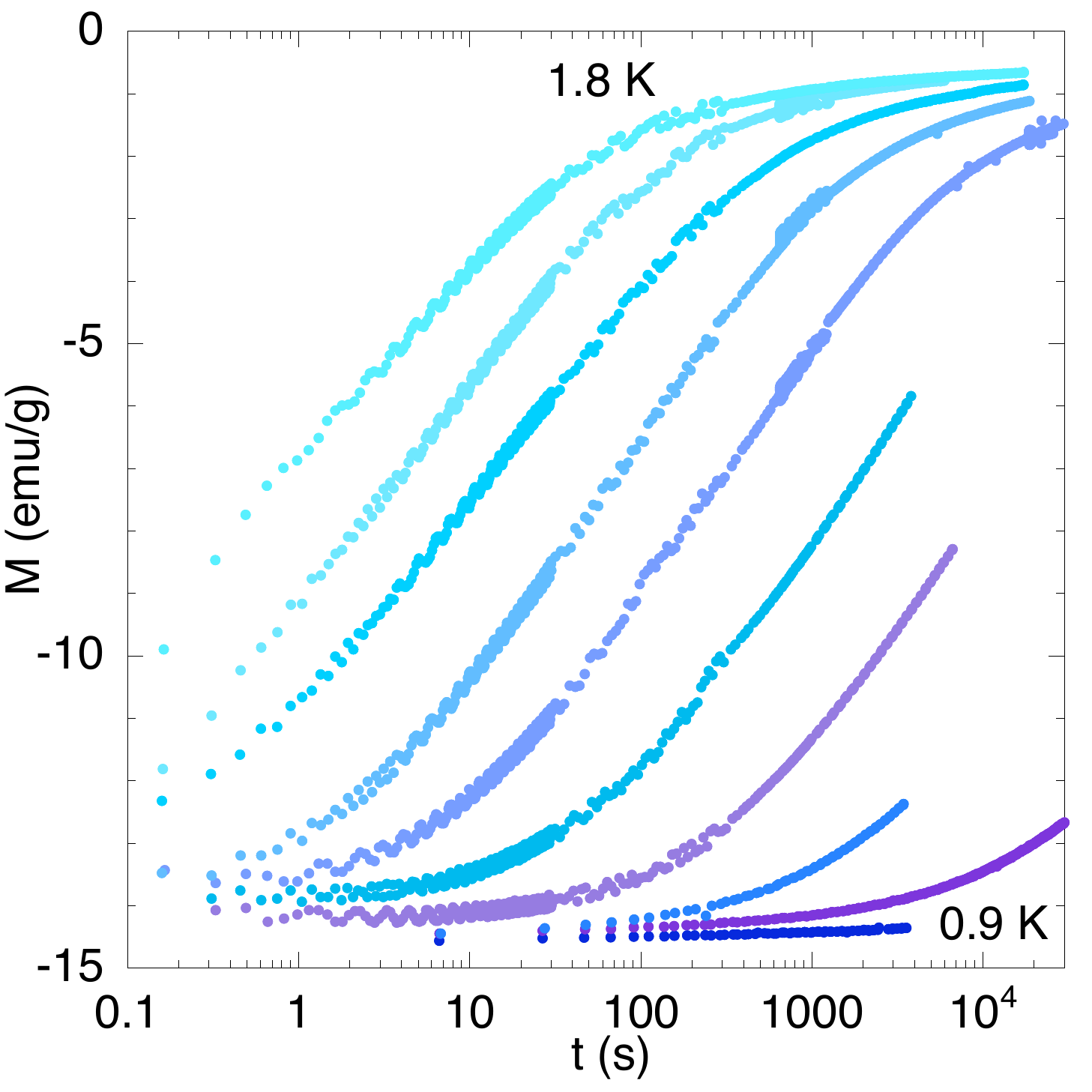}}
\caption{(color online) Magnetization $M$ vs time $t$ at several temperatures between 0.9 and 1.8 K. The sample is heated at about 2 K and cooled to the target temperature in an applied field of -3500 Oe. At $t=0$, the field is removed. }
\label{relax}
\end{figure}

\begin{figure}[h]
\centerline{\includegraphics[width=8cm]{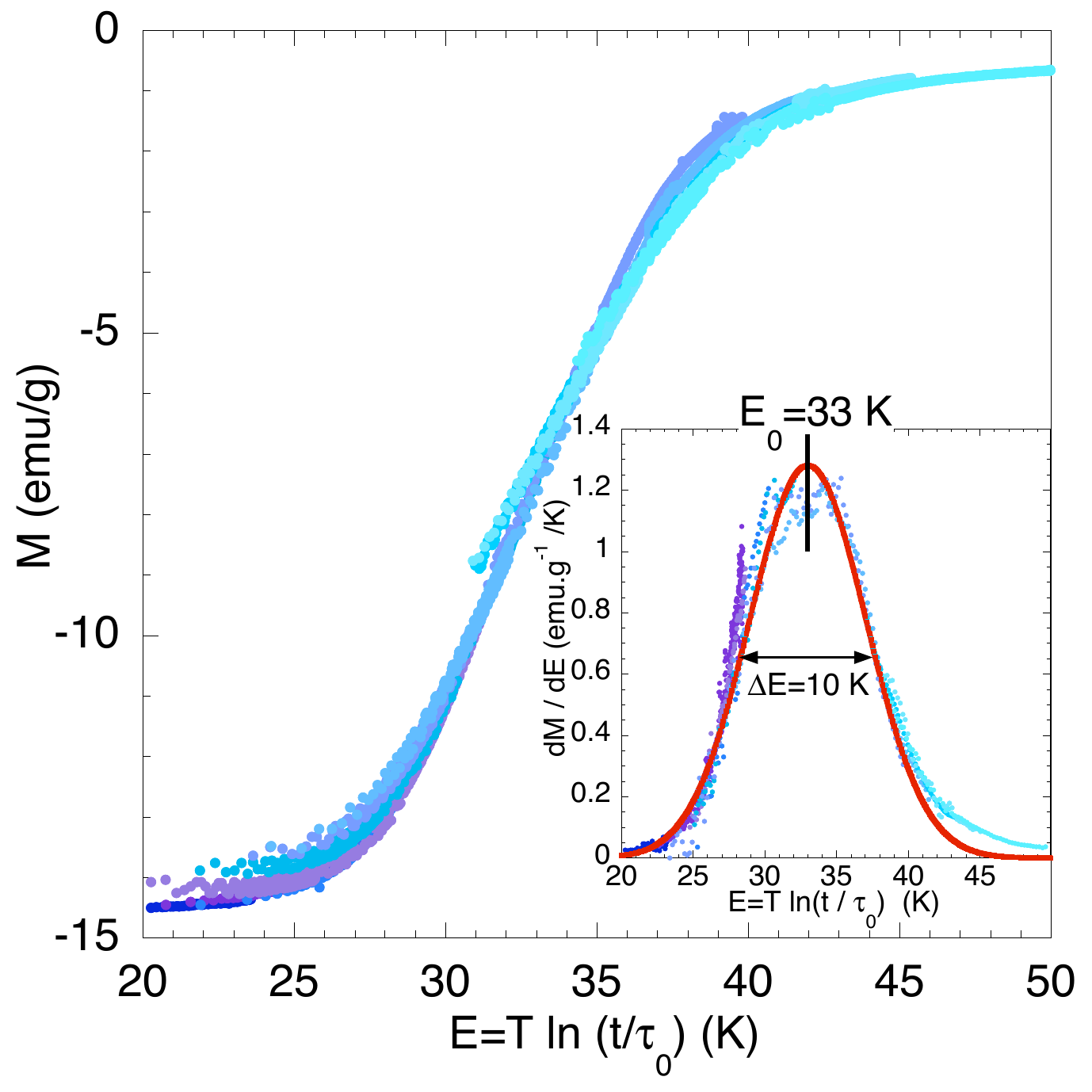}}
\caption{(color online) Magnetization $M$ vs the energy $E=T \ln (t/\tau_0)$, for 0.9 K$<T<1.8$ K. $\tau_0 =1.5 \times 10^{-8}$ s is chosen so that all the magnetization curves collapse. Inset: Derivative $dM/dE$ vs $E$ showing the barrier distribution (see text). The line is a Gaussian law centered on $E_0=33$ K with $\sigma^2$=16.5. }
\label{fig_MTlnt}
\end{figure}

The relaxation curves obtained in zero field are shown in Figure \ref{relax} for temperatures between 0.9 and 1.8 K. The relaxation is not purely exponential, and can be fitted with stretched exponential functions. This confirms the existence of a distribution of relaxation times. To investigate more precisely the barrier distribution, we followed the approach of Ref. \onlinecite{Prejean80} by plotting the measured magnetization as a function of a single variable, $E(T,t)=T \ln(t/\tau_0)$. All curves can be scaled using the parameter $\tau_0\approx1.5 \times 10^{-8}$ s (see Figure \ref{fig_MTlnt}), thus confirming the picture of a distribution of barriers crossed by thermal activation. A barrier distribution $P(E)$ can then be deduced from the derivative of the magnetization with respect to $E$, $dM/dE$, which is plotted on the inset of Figure \ref{fig_MTlnt}. 

The obtained barrier distribution is almost symmetric, but rather large: the full width at half maximum $\Delta E _{1/2}$ evaluated by the fit to a Gaussian function is about 10~K. Surprisingly, the barrier distribution is centered on $E_0\approx33$ K, that is to say about half of the energy barrier $E_b$ obtained from ac susceptibility measurements (in zero field). In addition, the value of $\tau_0$ is three orders of magnitude larger than the one deduced from ac measurements. These results indicate the existence of different relaxation processes for the two types of measurement. 
This is in strong contrast with other spin-chains on which the same analysis was made \cite{Lhotel08}, and where the obtained barrier energy was similar from ac susceptibility and relaxation of the magnetization.

\begin{figure}[t]
\centerline{\includegraphics[width=7.9cm]{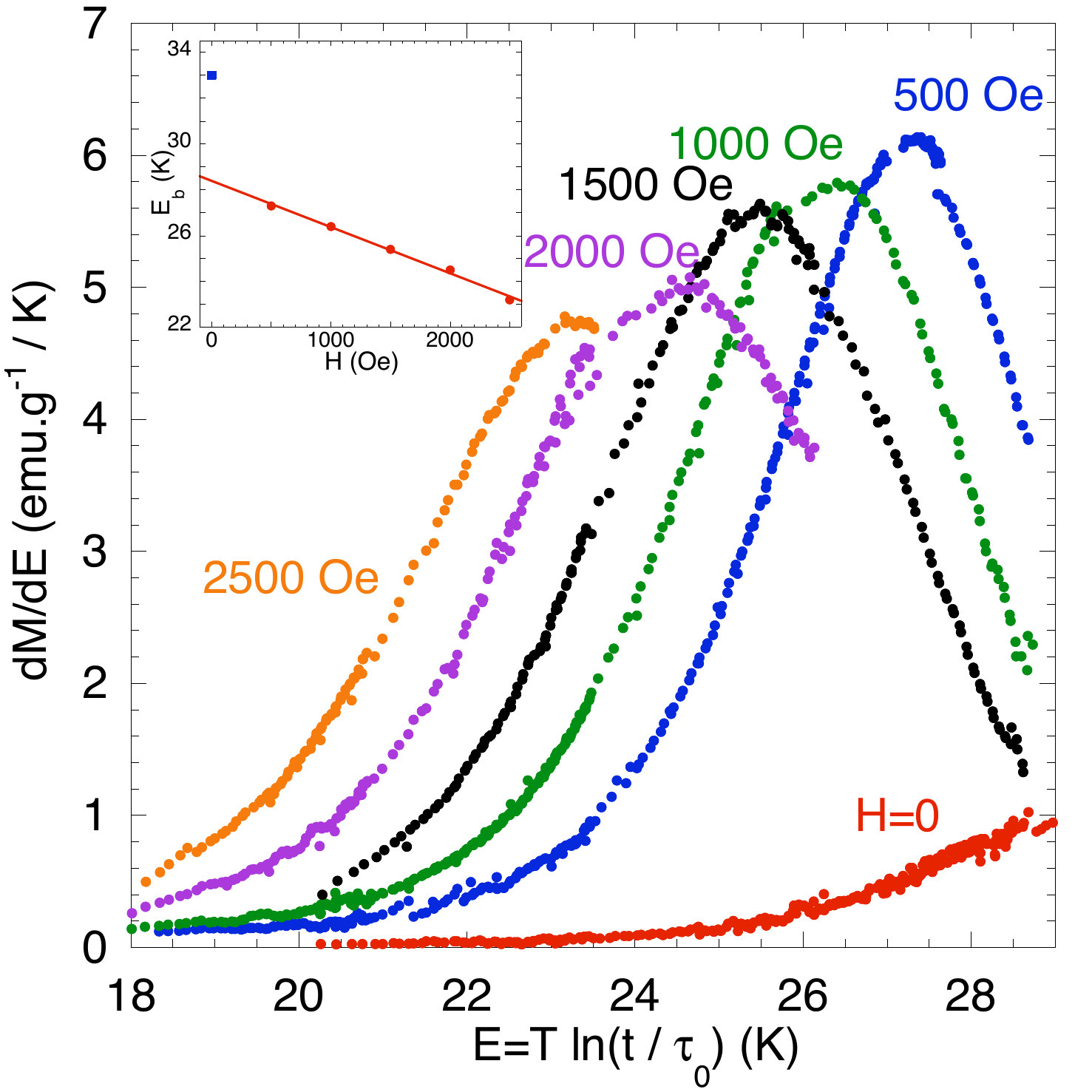}}
\caption{(color online) Derivative $dM/dE$ vs the energy $E=T \ln (t/\tau_0)$ with $\tau_0=1.5 \times 10^{-8}$ s, obtained for relaxation curves measured in applied fields of 500, 1000, 1500, 2000 and 2500 Oe between 0.8 and 1.2 K. The curve at $H=0$ (inset of Figure \ref{fig_MTlnt}) is shown for comparison. The inset gives the central barrier energy $E_b$ vs applied field $H$. The line is a fit for $H \neq 0$ to the equation: $E=-2.02 \times 10^{-3} H +28.4$.}
\label{fig_dMdEvsH}
\end{figure}

We have also performed measurements of the magnetization relaxation in non-zero applied fields, from 500 to 2500 Oe. The protocol described above was used for temperatures between 0.8 and 1.2 K. Note that for fields larger than 2000 Oe, and at high enough temperatures, the relaxation is so fast that avalanches occur. This phenomenon, which involves self-heating of the sample, prohibits access to the full barrier distribution at these fields. Nevertheless, the same scaling procedure was applied and led to the same $\tau_0$ ($=1.5 \times 10^{-8}$ s). Furthermore, the barrier distribution could be obtained (see Figure \ref{fig_dMdEvsH}). Two main features are clear from this figure. First, the distribution width is much smaller in applied field than in zero field. In the presence of an applied magnetic field, the full width at half-height $\Delta E _{1/2}$ varies between 3.5 and 4.5 K, a value to be compared with the 10 K obtained in zero applied field. Second, the mean energy barrier decreases when the field increases. The dependence is linear for non-zero applied fields (see inset of Figure \ref{fig_dMdEvsH}). This is consistent with the two-level system picture used for the analysis, in which the energy barrier would be reduced by the Zeeman energy: $E_H=E'_0- \mu_0 mH$, where $E'_0$ is the energy barrier in zero field, and $m$ is the magnetic moment parallel to the applied field. The obtained $E'_0$ value is smaller than the experimental $E_0$ ($E'_0=28.4$ K against $E_0\approx33$ K - see inset of Figure \ref{fig_dMdEvsH}), but within the barrier distribution measured at zero field. The slope $m$ is equal to 30 $\mu_B$. This might indicate that the energy barrier does not concern single spin flips but rather blocks of spins of effective moment $m$. 
In that sense, the larger distribution width, as well as the difference between $E_0$ and $E'_0$, might be explained by a more complex domain structure in zero field than in finite applied field for which a domain selection occurs (as shown by magnetization measurements).

\section{Discussion}
\label{discussion}
The complex magnetic structure and the diversity of the exchange paths in this cobaltous chain make the interpretation of the above results challenging. Nonetheless, the observed slow dynamics has a large signal, and is very clean and reproducible, suggesting that there should be a well-defined origin of this phenomenon. 
In the following, we examine the exchange paths and the relevant energy scales in the system, and present a possible scenario explaning the observations described in the previous section.

\subsection{Ingredients for the magnetic structure}
\label{discussion1}
\begin{figure}[h]
\centerline{\includegraphics[width=7.5cm]{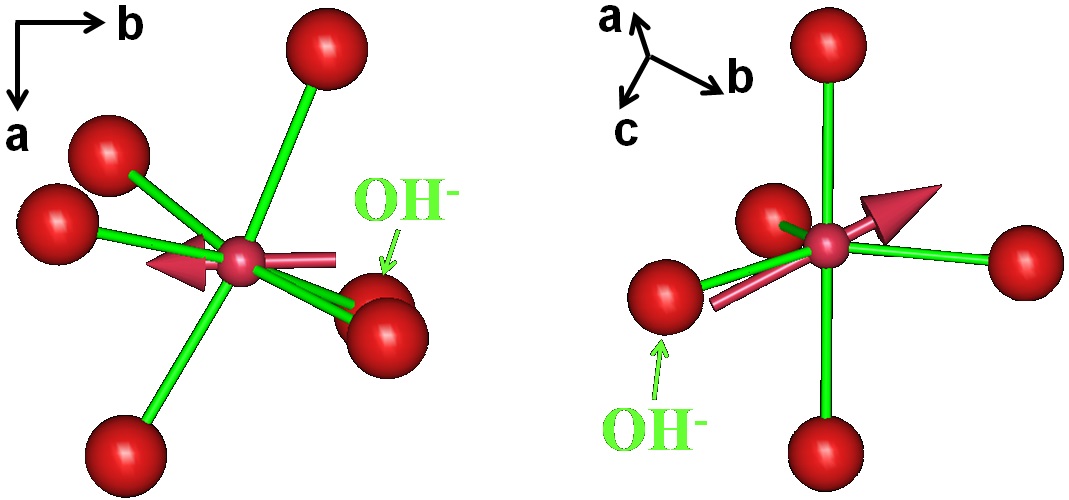}}
\caption{(color online) Schematic picture of the obtained moment orientation in their coordination polygone. Left: Co$^{\rm II}$1 site, right: Co$^{\rm II}$2 site. Both cations contain one hydroxide group among their ligand atoms.}
\label{fig_coordination}
\end{figure}

The magnetic structure results from the magnetic interactions and the local anisotropy of the Co$^{\rm II}$ ions. 
It is interesting to compare the anisotropy expected in the two types of environments adopted by the Co$^{\rm II}$ ions. 
As mentioned in Section \ref{struct}, the Co$^{\rm II}$1 coordination consists of distorted octahedra, whereas the Co$^{\rm II}$2 ion is 5-coordinated in a trigonal bipyramid. In the former, the magnetic moment is almost in the basal plane, like already observed in other magnetic structures involving Co$^{\rm II}$ ions in similar environments.\cite{Kinast10,Viola03}.  The J1 exchange coupling (see below) could be responsible for the proposed orientation, thus implying that this coupling is stronger than the in-plane anisotropy of the Co$^{\rm II}1$ ions. Nevertheless, our neutron data cannot exclude that the magnetic moments lie exactly in the basal plane. 
In the latter, Co$^{\rm II}$2, the anisotropy can be stronger \cite{Foglio02,Robertson11} than in the octahedral coordination: it has been observed to be even larger when the Co$^{\rm II}$ ion is out of the basal plane, which is the case here (see Figure \ref{fig_coordination}), and to be of order 25~K \cite{Jurca11}.

We examine now the exchange paths that appear to be relevant for the understanding of the intrachain magnetic structure. Their estimation is deduced from both the Goodenough-Kanamori rules\cite{Good63}, and the values reported in cobalt-based magnetic systems with similar exchange networks \cite{Kurmoo09}. 
They are summarized in both Table \ref{tableJ} and Figure \ref{Fig.2}. 
The strongest magnetic coupling J1 is antiferromagnetic and induces the collinear antiferromagnetic arrangement of the Co$\textsuperscript{II}1$ sublattice. The J2 (weak ferromagnetic) and J3 (weak antiferromagnetic) interactions determine the relative orientation between the Co$^{\rm II}$1 and Co$^{\rm II}$2 sublattices. The strong canting of the Co$^{\rm II}2$ ion with respect to the Co$^{\rm II}$1 ion demonstrates the importance of the single-ion anisotropy of this ion. In that sense, given the anisotropy of the ions, the experimentally deduced magnetic structure satisfies the three interactions J1, J2 and J3. The J4 coupling, which is expected to be weakly antiferromagnetic, is frustrated in the obtained magnetic structure, thus suggesting a smaller coupling compared with J2 and J3.

\begin{table}[h]
\begin{ruledtabular}
\begin{tabular}{*{5}{c}}
Exchange& Exchange & Sharing & Exchange & Estimated \\
constant &path&type&type &value \\ \hline
J1 & Co1a-OH-Co1b & corner & AF & [-20,-40] K \\ 
J2 & Co1-O-Co2& edge & F & [10, 30] K \\
J3 & Co1-OH-Co2& corner & AF & [-15,-25] K \\
J4 & Co1-O-Co2 & corner & AF & [-15,-25] K \\ 
\end{tabular}
\end{ruledtabular}
\caption{Exchange paths and their estimation following Ref. \onlinecite{Good63, Kurmoo09}.}
\label{tableJ}
\end{table}

Given the structural characteristics of {\CoChain} presented in Section \ref{struct}, the 3D magnetic LRO in this material can only be explained by interchain dipolar interactions. The energy of the dipolar interaction between two spins $\textbf{S}_{i}$ and $\textbf{S}_{j}$ separated by a distance $\textbf{r}_{ij}$ is

\begin{equation*}
E_{\rm dip} = \mu_{B}^2g^2\frac{1}{r_{ij}^3}\left[\textbf{S}_{i}.\textbf{S}_{j}-\frac{3(\textbf{S}_{i}.\textbf{r}_{ij})(\textbf{S}_{j}\textbf{r}_{ij})}{r_{ij}^2}\right]
\label{dipolar}.
\end{equation*}

In the present case, the interchain dipolar interaction is typically a few millikelvins, and implies a ratio of about 10$^4$ between the intrachain and interchain interactions. 
In such a system with a strong 1D character, the critical temperature describing LRO originates from the combination of the strong intrachain correlations, resulting in spin blocks exhibiting a 'super-spin' (c.f. uncompensated component of the magnetic structure), and the interchain interactions between these super-spins \cite{Drillon98,Ostrovsky01,Lhotel07}. In the present material, this picture is most readily applied along the crystallographic \textbf{c}-axis which corresponds to the shortest interchain distance. In that direction the dipolar interaction stabilizes an interchain ferromagnetic arrangement between the intrachain super-spins. However, the complexity of the magnetic structure makes this description much more complicated for the ${\bf a}$ direction. The reproduction of the magnetic motif observed in this direction is not favourable when considering the orientation of the super-spins emerging from the ferromagnetic component.
The dipolar interactions between individual spins have thus to be considered along \textbf{a}-axis. Preliminary calculations of the dipolar energy suggest that the obtained orientation is more energetically favourable with respect to the dipolar interaction than its 180$^{\circ}$ counterpart.

\subsection{Origin of the slow dynamics}
In LRO spin chains, very strong correlations exist within the chains, while the spin chains are weakly coupled to each other. The result is that when the temperature is larger than the interchain interactions (of a few millikelvins in our case), the dynamics are governed by intrachain dynamics, which can be described in terms of domain-walls motion. This was observed in several ordered spin chains \cite{Lhotel08, Lhotel06, Coulon09}: dependent on the ratio between the local anisotropy and exchange interactions, the dynamics can be described in terms of single-chain magnets or motion of extended domain-walls.

In \CoChain, the presence of two Co$^{\rm II}$ sites with a different single-ion anisotropy, combined with the existence of several exchange couplings, makes the description in terms of domain-walls complicated. To get a better understanding, we return to the striking result of the analysis of the dynamics: the ac susceptibility (i.e zero-field) dynamics is accounted for by an energy barrier of $E_b\approx64$ K over a large range of temperature, whereas the relaxation of the magnetization (i.e. from a finite magnetization state) can be described by a barrier of half value $E_0\approx33$ K. To go further, we will have to keep in mind that the magnetic structure is only canted by the field applied during measurements of the magnetization relaxation (see Section \ref{neut_field}).

The zero-field dynamics measured by ac-susceptibility can be ascribed to the motion of domain-walls along the chains. Due to the strong anisotropy, we can expect that these domain-walls are relatively narrow. Then the observed energy barrier would be due to the cost in energy (exchange and anisotropy) to move the wall at the domain boundary. The ranges of the exchange constants and anisotropy (see Section \ref{discussion1}) is consistent with the obtained energy barrier. 

We recall that (i) magnetization relaxation measurements were done after a field of -3500 Oe was applied, and that (ii) the sample being a powder, the field acts on the three directions of the crystal. Moreover, as already mentioned, a fast increase of the magnetization is observed when a field is applied. This can be attributed to a selection of domains with respect to the resulting spin component in the ferromagnetic-like direction, the $\bf c$ direction. In the other directions, or at larger fields, the magnetization increases smoothly by spin canting. Because of the strong anisotropy of the Co$^{\rm II}$2 ions, this induced magnetization can be assigned to the Co$^{\rm II}$1 ions. 
This canting is effective only along the directions {\bf a} and {\bf c}, since the  Co$^{\rm II}$1 ions are oriented along the {\bf b}-axis. 

When the field is removed, the return to equilibrium, that is to say, the relaxation of the magnetization, will consist of two processes: the reorientation of the Co$^{\rm II}$1 ions and the creation of domains to recover the zero magnetization.   
It is only the second process which needs to overcome a barrier by nucleation of domain-walls inside the chains. The obtained $E_0$ value should be associated with this nucleation. 

The question that immediately arises is why $E_0$ is half of $E_b$. A possible reason could be that the nucleation occurs at the boundary of a chain (or near a defect), where it is more energetically favourable, thus creating only one domain-wall. On the contrary, the dynamics probed by ac susceptibility concern existing domains where the two boundaries of the domains are moving in response to the ac-field (because the magnetization has to remain zero). The obtained characteristic relaxation time $\tau_0$ is three orders of magnitude larger in magnetization measurements than in ac susceptibility measurements, suggesting that the nucleation process is much slower. 

To get a deeper insight in these dynamics, and especially the field induced dynamics, further measurements on single crystals are needed. These would allow us to determine if an anisotropy exists in the dynamics (which should be the case in our scenario), and to provide a more quantitative analysis of these observations. 

\section{Conclusions}
\label{conclusions}
In conclusion, we have determined the magnetic structure of \CoChain. It is made of spin-chains involving two sublattices: one is purely antiferromagnetic, the second gives rise to a ferromagnetic component along the \textbf{c}-axis. Due to the magnetic propagation vector being \textbf{\textsl{k}}=(0,0,0), the magnetic 'motif' is reproduced in the neighboring unit cells forming the three dimensional lattice. The magnetic structure at low temperature is not modified upon the application of a magnetic field, at least up to 2~T. Only canting within the chains is observed. Furthermore, the applied field reinforces the interchain correlations resulting in a higher ordering temperature which increases up to 13~K at 2~T (to be compared to the transition temperature $T_t=5.4$~K).
At very low temperature, in the ordered phase, slow dynamics are observed in a large temperature/frequency range which are ascribed to dynamics of domains inside the chains. Interestingly, these dynamics are extremely sensitive to the presence of a magnetic field.

Finally, we notice that the nature of the present material, comprising inorganic chains interconnected through organic linkers, is reminiscent of other metal-organic frameworks that have shown high intrinsic flexibility in response to high pressures \cite{Ortiz12, Gagnon13, Lapidus13} or to the absorption of guest molecules \cite{Serre02}. This point is particularly interesting as it might be possible to tune the organization of the chains relative to each other and thus the magnetic ground state. For this purpose, powder X-ray diffraction experiments under back pressure of CO$_2$ (up to 100 bars) and H$_2$ (up to 200 bars) were performed. However, the crystal structure of \CoChain did not seem to be affected. Experiments under higher pressure, using a diamond anvil cell, will be the scope of future experiments.

\acknowledgments 
C. Paulsen is warmly acknowledged for the use of his magnetometers. The authors thank the Institut Laue Langevin, Grenoble, France for provision of the research facilities. T. Hansen and A. Daramsy are acknowledged for their assistance during the D20 experiments. R.S. thanks MESR for his Ph.D. funding and J. S. White for reading of the article. We acknowledge the support of the European Community - Research Infrastructures under the FP7 Capacities Specific Programme, MICROKELVIN project number 228464.


\begin{thebibliography}{}
\bibitem{footnote1} Ideally, the sample should be heated over $T_t=5.4$ K, but the working conditions of the dilution fridge prevents from heating to such a high temperature.
\end{thebibliography}
\end{document}